\documentclass[a4paper,12pt]{article}
\usepackage{amsmath,amssymb,epsf}
\usepackage{amssymb,amsmath,epsfig,graphicx}
\usepackage{ifthen,theorem,color}

\begin{document}

\begin{center}
{\bf Quantum Ground State Energies for Very Flat Potentials}\\
Rodney O. Weber\\
School of Mathematics and Applied Statistics\\
University of Wollongong\\
Northfields Avenue, Wollongong, NSW 2522\\
Australia\\
rweber@uow.edu.au\\
\end{center}

{\bf Abstract}

An infinite sequence of potential well functions is considered. A trial wavefunction is 
used with the Schr$\ddot{\text{o}}$dinger equation to obtain an approximate ground state 
energy for each potential well function. We obtain an expression that is exactly correct 
for the harmonic potential, has an intuitively correct form for all of our potential well
functions and can be understood to be a partitioning of the ground state energy into
two parts, one kinetic and one potential. 

{\bf 1. Introduction}

The quantum theory of a particle in a potential has been much studied over the last hundred years.
The infinite square well and harmonic potentials are two of the best known examples where a complete
analytic solution is possible and all of the energy levels (eigenvalues) and wave functions 
(eigenfunctions) can be completely determined; e.g. Davies (1984).
Most potential functions do not admit exact analytic solution, but fortunately there are useful
analytic methods that can be used to approximately solve for the eigenvalues and eigenfunctions.
In the current work we will calculate approximate results for the ground state energy level 
for a sequence of potentials, 
beginning with the harmonic potential, each one getting progressively flatter
and finally the same as the infinite square well. We will use trial wavefunctions whose
functional dependence will reflect the progressive changes in the potentials. These trial
wavefunctions are in fact generalised Gaussians and their shape is indicative of our
intuitive expectation that as the potentials become steeper, the wavefunction should
be progressively more localised. The functional form of the calculated ground state energies
will be seen to be exactly as one would expect, although the approximate numerical values
multiplying the functional form will be seen to be most accurate in the case of the harmonic
potential (where it is exactly correct) and gradually becoming less accurate as the 
polynomial order of the potentials increases. 
The key observation from this is that the ground state energy is partitioned between two
contributions, similar to a kinetic energy part and a potential energy part and that   
the ground state energy has a predictable, useful, functional dependence.

{\bf 2. Square Well Sequence}

Consider the potential function $V(x)= \mu (x/a)^N$ where $\mu$ and $a$ are constants 
and $N$ is a positive, even, natural number. 
$N=2$ corresponds to the familiar harmonic potential, $N=4$ is purely quartic and
usually referred to as the anharmonic potential, $N=6$ is the sextic potential, $N=8$ is the octic
potential and so on. For $N$ extremely large the sequence becomes identical to the infinite
square well of width $2a$. Note that we have written these potentials with $x/a$ to ensure that
each potential in our sequence can be neatly overlaid on one graph, illustrating how the
sequence of potentials becomes successively flatter and how it (the sequence) approaches
the infinite square well. The explicit appearance of $\mu$ allows us to easily compare
the strength of our potential with the slightly more familiar forms and in the case of 
the infinite square well one can set $\mu=1$ without loss of generality.  

A fact that we will make use of later is that we can use the absolute value of $x/a$ 
and raise this to the power of $N$ and then we need not be restricted to $N$ being an 
even, natural number; $N$ can then be any real number greater than or equal to $2$ 
and all of the conclusions that we shall elucidate below will still be correct. 

{\bf 3. Ground States: Exact and Estimated Known Results}

Firstly we note that for the infinite square well, of width $2a$, the ground state energy is 
$E = \frac{ \pi^{2} \hbar^{2}}{8ma^2}$ (we have set $\mu=1$ as suggested above). 

For the harmonic oscillator case, $N=2$, we can make use of the exact result for the energy
eigenvalues. In particular the ground state energy is $E= \frac{1}{2} \hbar \omega$ with frequency $\omega = \sqrt(\frac{2 \mu}{m a^2})$, whereupon we write the ground state 
energy as 
$E = \frac{1}{\sqrt{2}}\left(\frac{\hbar^{2}}{ma^{2}}\right)^{\frac{1}{2}}\mu^{\frac{1}{2}}$.

For the purely quartic potential a complete analytic solution is not possible. However,
numerous investigations have reported excellent approximate results. Bender and Wu (1969)
and (1973) elaborated on the use of perturbation methods and discovered the singularities
that are responsible for the divergence of the perturbation series. Janke and Kleinert (1995)
employed variational perturbation theory to give an extremely accurate result for eigenvalues. 
In particular, they found that the ground state energy for the purely quartic case is 
$$
E = 0.667986259155777.....(\lambda/4)^{1/3}; 
$$
given here to only 15 significant digits, but correct upto 23 digits and in agreement with 
the most accurate numerical result known at the time, which was calculated and reported by
Vinette and Cizek (1991). The potential strength they used, $\lambda/4$, is equivalent to our $\mu/a^4$.
More recently, Liverts {\it et al} (2006) used an iterative technique to
determine the eigenvalues and the eigenfunctions and Frasca (2007) used a Wigner-Kirkwood
expansion to determine the eigenvalues and make some interesting observations about strongly 
perturbed quantum systems and semiclassical systems. Either way, the result for the ground
state energy given in the equation above for the purely quartic case is the key result 
for our present purposes.

The sextic potential we write as $\mu (x/a)^6$, the octic potential we write as $\mu (x/a)^8$, 
etc. Sextic and octic potentials were examined by Weniger et al (1993) using extremely
sophisticated methods including Pade approximations to rearrange series, but higher powers
have to the best of our knowledge not been greatly studied analytically and
there are no exact values for the ground state energy.

{\bf 4. Ground State Energies: Approximate Method}

We now wish to show an approximate method, motivated by the anticipated asymptotic behaviour
of the wavefunctions,  that will allow us to estimate the ground state energy for all
of these potential functions. We begin by assuming that a Gaussian like function will
be a reasonable representation of the asymptotic behaviour of the wave functions for
large, negative or positive, $x$. That is,
$$
\psi=Ae^{-\alpha\left|\frac{x}{a}\right|^{\beta}},
$$
with $A$ a normalisation constant and $\alpha$ and $\beta$ contstants that we are yet to determine. The normalisation condition $\int_{-\infty}^{\infty}\psi^{2}dx=1$ 
results in an equation for $A$, namely
$$
A^{2}\int_{-\infty}^{\infty}e^{-2\alpha\left|\frac{x}{a}\right|^{\beta}}dx=1.
$$
Notice that we have used $|x/a|$ anticipating that $\beta$ will not necessarily be an even, 
natural number in the calculations that will follow and hoping to still ensure convergent integrals. All of our subsequent results will hold for any $\beta$ greater than or equal to two.

We will use this asymptotic approximate form as a `trial solution' for the wavefunction in the 
Schr$\ddot{\text{o}}$dinger equation
$$
-\frac{\hbar^{2}}{2m}\frac{d^{2}\psi}{dx^{2}}+\mu\left(\frac{x}{a}\right)^{N}\psi=E\psi,
$$
and then we will multiply the Schr$\ddot{\text{o}}$dinger 
equation by the complex conjugate of $\psi$
and integrate from $-\infty$ to $+\infty$.
The resultant equation, with our approximate, or trial, $\psi$ substituted in, becomes
$$
\int_{-\infty}^{\infty}\left(-\frac{\hbar^{2}}{2m}\frac{\alpha^{2}\beta^{2}}{a^{2}}\left|\frac{x}{a}\right|^{2\beta-2}+\frac{\hbar^{2}}{2m}\frac{\alpha\beta(\beta-1)}{a^{2}}\left|\frac{x}{a}\right|^{\beta-2}+\mu\left(\frac{x}{a}\right)^{N}\right)A^{2}e^{-2\alpha\left|\frac{x}{a}\right|^{\beta}}dx 
$$

$$
= \int_{-\infty}^{\infty}E~A^{2}e^{-2\alpha\left|\frac{x}{a}\right|^{\beta}}dx. \hskip6cm (1)
$$
To arrive at this form we have taken the second derivative of $\psi$ and noted that for our 
approximate, or trial, function the complex conjugate is identical to the function itself.

It seems as though the normalisation constant $A$ cancels out from both sides of this
equation, however the presence of the integral on the right hand side, multiplying $E$,
is in fact the definition of $A^2$.

In order to proceed further we now consider our options for the constants $\alpha$ and $\beta$,
introduced in the definition of our trial function for $\psi$.
As the left hand side consists of three terms, each with a prima facie different power of $x$,
we consider equating the highest power containing $\beta$ with $N$.
This suggests $2 \beta - 2 = N$ or $\beta = (N+2)/2$. 
We can also then equate the coefficients of these two powers of $x$ and use the fact that
they have opposite signs to achieve a cancellation.
There results an equation for $\mu$ in terms of $\alpha$
$$
\mu = \hbar^2 \alpha^2 \beta^2 /2 m a^2.
$$
This can be considered to be a way to decide upon suitable values of $\alpha$ and $\beta$
that ensure that our trial function $\psi$ has the correct asymptotic behaviour.
In all, this means that have now determined all three constants in our trial function.
We can then simplify and re-arrange the Schr$\ddot{\text{o}}$dinger equation 
(eq.1) to arrive at an estimate for the energy, namely

$$
E=\left(\frac{\hbar^{2}}{2ma^{2}}\right)\alpha\beta(\beta-1)\frac{\int_{-\infty}^{\infty}\left|\frac{x}{a}\right|^{\beta-2}e^{-2\alpha\left|\frac{x}{a}\right|^{\beta}}dx}{\int_{-\infty}^{\infty}e^{-2\alpha\left|\frac{x}{a}\right|^{\beta}}dx}\\
$$
and if we introduce a non-dimensional $z = x/a$ we finally obtain
$$
E=\left(\frac{\hbar^{2}}{2ma^{2}}\right)^{1-\frac{1}{\beta}}\mu^{\frac{1}{\beta}}\left(\frac{\beta(\beta-1)}{\beta^{2/\beta}}\right)\frac{\int_{-\infty}^{\infty}\left|z\right|^{\beta-2}e^{-2\alpha\left|z\right|^{\beta}}dx}{\int_{-\infty}^{\infty}e^{-2\alpha\left|z\right|^{\beta}}dx}, \hskip1cm (2)
$$
with $\beta=\frac{1}{2}(N+2)$ and if $N=2k\hspace*{0.1in}\Rightarrow\hspace*{0.1in}\beta=k+1$.

This final form for the ground state energy is our main result. It provides a
compact statement of the energy for any $N$ greater than or equal to $2$. Clearly to arrive
at a numerical answer one must still carry out the integrations in the numerator
and denominator and as far as we are able to establish these must be done numerically
for each value of $\beta$. So to show actual numerical results, the procedure is 
to choose a value for $N$ and from it determine $\beta$ and then perform the required
numerical integrations. For the harmonic case $N=2$, whence $\beta=2$, and the result
given in eq.(2) immediately simplifies to the well known and exact result as expected
and given earlier in section 3. 
For all other values of $N$ the integrals need to be computed as indicated.  
In the following table we list several results for the ground state energy.
The numerical integrals have been performed using the statistical software package R.
Note that we have chosen $N$ to be even in each case as quartic, sextic, etc. potentials
are the ones usually considered to be of interest. However, as alluded to earlier,
we can use the absolute value of $x/a$ and then it is possible to consider potentials
with odd values for $N$ such as cubic, etc. In fact the integrals from $-\infty$ to $+\infty$
can be doubled up and just evaluated from $0$ to $+\infty$ to make calculations
slightly simpler and then we can even dispense with the absolute value
in the case of odd values for $N$ and/or $\beta$.

\begin{table}[!htp]
\centering
\caption{Results}
\begin{tabular}{c|l}
\hline
Potential & Ground State Energy\\\hline
$\mu\left(\frac{x}{a}\right)^{2}$ & $\frac{1}{\sqrt{2}}\left(\frac{\hbar^{2}}{ma^{2}}\right)^{\frac{1}{2}}\mu^{\frac{1}{2}}$\hspace*{0.2in} `exact'\\
$\mu\left(\frac{x}{a}\right)^{4}$ & $0.7290111\left(\frac{\hbar^{2}}{ma^{2}}\right)^{\frac{2}{3}}\mu^{\frac{1}{3}}$\\
$\mu\left(\frac{x}{a}\right)^{6}$ & $0.8526415\left(\frac{\hbar^{2}}{ma^{2}}\right)^{\frac{3}{4}}\mu^{\frac{1}{4}}$\\
$\mu\left(\frac{x}{a}\right)^{8}$ & $1.009593\left(\frac{\hbar^{2}}{ma^{2}}\right)^{\frac{4}{5}}\mu^{\frac{1}{5}}$\\
$\mu\left(\frac{x}{a}\right)^{2k}$ & $(number)\left(\frac{\hbar^{2}}{ma^{2}}\right)^{\frac{k}{k+1}}\mu^{\frac{1}{k+1}}$\\
Infinite Square Well & $\frac{\pi^{2}}{8}\left(\frac{\hbar^{2}}{ma^{2}}\right)$\hspace*{0.2in} `exact'\\
\hline
\end{tabular}
\end{table}

The key observation from considering each of the above potential well functions is that
there is a particular combination of the constants in the problems that always
appears in the equation for the eigenvalues and hence the ground state energy; namely
powers of $\hbar^2/ma^2$ and powers of $\mu$. Checking the dimensions of the result
is easy as $\hbar^2/ma^2$ has the dimensions of energy, as does $\mu$. And the two powers
always add to one. Hence, it is as though the ground state energy is partitioned between
the two contributions; one kinetic energy and the other potential energy.

{\bf 5. Discussion}

We begin by observing that the estimates for the ground state energy seem quite good
for the quartic, sextic and octic potentials. Unfortunately, as $N$ gets larger (and
hence $k$ because $N=2k$ and $\beta$ because $\beta=k+1$) the numerical values 
from the integrals in the ground state energy estimate (eq.2) appears to become increasingly
inaccurate. It would have been satisfying if the numerical value approached $\pi^2/8$,
the value for the infinite square well, but this does not seem to be the case. 
Investigation of the asymptotic behaviour of the integrals for large $\beta$ is 
currently underway, but is unlikely to offer any additional insight. 

Additionally, we should here emphasize that our trial wavefuntion is generally not
the actual wavefunction for the ground state. We could use a variational method to
determine how close the trial functions are to the actual ground state wavefunctions,
but our infinite sequence of potentials means we are considering an infinite number
of trial wavefunctions. Instead a simple graphical procedure allows almost the
same insight. Plotting the trial wavefunctions makes it abundantly clear that as
the power $N$ increases, the trial wavefunctions become very localised to the 
centre of the $x$ region and in the limit as $N$ goes to infinity the trial 
wavefunction will never be close to the known infinite square well wavefunction.   

On the other hand, the partitioning of the ground state energy estimate between
contributions $\hbar^2/ma^2$ and $\mu$ seems to be correctly given and is 
intuitively correct, beginning with the well known exact result for the harmonic
potential, progressing through the known result for the pure quartic potential 
and asymptotically approaching the well known exact result for the infinite 
square well. Along the way this illuminates the hitherto unknown, or at least
little reported, functional behaviour for the cases of the sextic, octic, etc. potentials.
Furthermore, with the use of the absolute value of $x/a$, the present method
allows the calculation of estimates for the ground state energy of potential
functions based on odd powers of $x$ such as the cubic and quintic.

{\bf 6. Application and Further Work}

The quantum ground states for these very flat potentials may have application in quantum
chemistry; for example Germanium Carbide (Zingsheim, 2017) where recent work has conclusively
established a T-Shaped structure for the ground state. There may be scope to model the 
bonds with a mix of quartic and higher order potentials as these are just an
approximation to very flat potential wells that are found in the real world.
There could also be implications for zero point energy with the current observations 
offering alternatives for the minimum average energy of quantum fields in the vacuum. 
Finally, the development of a set of trial wave functions that are even better at
capturing the expected localised behaviour as the potentials steepen is being considered. 
It may be that this can only be carried out computationally, but several avenues are being
actively pursued. If successful, this would be likely to result in better numerical values 
for the approximate ground state energies.

\newpage

{\bf References}\\
C.M.Bender and Wu (1969) Phys.Rev.D 184 p.1231.\\
C.M.Bender (1973) Phys.Rev.D7 p.1620.\\
P.C.W. Davies (1984) Quantum Mechanics. Chapman \& Hall (reprinted 1990)\\
M.Frasca (2007) Proc.Royal Soc.A. 463 pp.2195.\\
W. Janke and H.Kleinert (1995) Phys.Rev.Lett. 75 p.2787.\\
E.Z.Liverts, V.B.Mandelzweig and F.Tabakin (2006) J.Math.Phys. 47 p.62109\\
F.Vinette and J.Cizek (1991) J.Math.Phys. 32 p.3392.\\
E.J.Weniger, J.Cizek and F.Vinette (1993) J.Math.Phys. 34 p.571.\\
O.Zingsheim, M.Martin-Drumel, S.Thorwirth, S.Schlemmer, C.A.Gottlieb, J.Gauss and
M.C.McCarthy (2017) J.Phys.Chem.Lett. 16 p.3776.\\

{\bf Acknowledgments}\\
The author would like to thank Dr.J.S.Chapman for assistance with LaTeX and R.

{\bf Authors' Contributions}\\
This paper is solely the work of the single author.

{\bf Competing Interests}\\
The author declares no competing interests.

{\bf Data Accessibility}\\
This paper has no data.

{\bf Ethics Statement}\\
There were no ethical considerations required for this work.

{\bf Funding Statement}\\
This work was done without funding.

\end{document}